\documentclass[twocolumn,prl,aps,tightenlines]{revtex4}
\usepackage{amsfonts}
\usepackage{amssymb}
\usepackage{epsfig}
\usepackage{amsbsy}
\usepackage{amsmath}
\usepackage{graphicx}
\usepackage{graphicx}
\usepackage{dcolumn}
\usepackage{bm}
\usepackage{color}
\usepackage{amsxtra}

\newcommand{\be}{\begin{equation}}
\newcommand{\ee}{\end{equation}}
\newcommand{\bea}{\begin{eqnarray}}
\newcommand{\eea}{\end{eqnarray}}
\newcommand{\bes}{\begin{subequations}\begin{eqnarray}}
\newcommand{\ees}{\end{eqnarray}\end{subequations}}

\begin{document}

\title{A variational approach for many-body systems at finite temperature}
\date{\today }
\author{Tao Shi$^{1,2}$, Eugene Demler$^{3}$, and J. Ignacio Cirac$^{4}$}
\affiliation{$^{1}$ Institute of Theoretical Physics, Chinese Academy of Sciences, P.O.
Box 2735, Beijing 100190, China \\
$^{2}$ CAS Center for Excellence in Topological Quantum Computation,
University of Chinese Academy of Sciences, Beijing 100049, China \\
$^{3}$ Department of Physics, Harvard University, 17 Oxford st., Cambridge,
MA 02138 \\
$^{4}$ Max-Planck-Institut f\"{u}r Quantenoptik, Hans-Kopfermann-Strasse. 1,
85748 Garching, Germany}
\date{\today }

\begin{abstract}
We introduce a non-linear differential flow equation for density matrices
that provides a monotonic decrease of the free energy and reaches a fixed
point at the Gibbs thermal state. We use this equation to build a
variational approach for analyzing equilibrium states of many-body systems
and demonstrate that it can be applied to a broad class of states, including
all bosonic and fermionic Gaussian states, as well as their generalizations
obtained by unitary transformations, such as polaron transformations, in
electron-phonon systems. We benchmark this method with a BCS lattice
Hamiltonian and apply it to the Holstein model in two dimensions. For
the latter, our approach reproduces the transition between the BCS pairing
regime at weak interactions and the polaronic regime at stronger
interactions, displaying phase separation between superconducting and
charge-density wave phases.
\end{abstract}

\maketitle
\affiliation{CAS}



Variational methods constitute powerful techniques to describe certain
many-body quantum systems \cite{BCS} in thermal equilibrium. Their
underlying principle is based on the fact that the free energy attains the
minimum value for the Gibbs state \cite{Statisticalphys}, which describes
the system in equilibrium. The success of such methods crucially depends on
both, the choice of the family of states and the technique used to carry out
the minimization. The choice of the variational states, on the one hand, has
to be broad enough to faithfully represent the physical behavior of the
system under study and, on the other, to be amenable to an efficient
computation of the observables of interest. The minimization procedure has
to be efficient as well, and avoid getting stuck in local minima.

For a system at zero temperature, a particularly successful minimization
procedure consists of using the time-dependent variational principle (TDVP)
in imaginary time. This approach, which we will call imaginary-time
variational method (ITVM), is based on the fact that given any state, $\Phi
(0)$, if we compute $\Phi (\tau )$ according to
\begin{equation}
d_{\tau }|\Phi \rangle =-[H-h(\tau )]|\Phi \rangle  \label{Htau}
\end{equation}%
where $H$ is the system Hamiltonian, then $h(\tau )=\langle \Phi |H|\Phi
\rangle $ decreases monotonically with $\tau $. The variational procedure
consists of projecting (\ref{Htau}) onto the tangent plane of the manifold
defined by the corresponding family of states \cite{Kraus,Tao}, so that one
obtains a set of (non-linear) differential equations for the variational
parameters, $\xi (\tau )$. The solution of such equations in the limit $\tau
\rightarrow \infty $ yields the variational state that minimizes the energy
\cite{footnote}. The ITVM has also been applied to systems at finite
temperature, $T$, by evolving a completely mixed state (or, more precisely,
its purification, see Ref. \cite{QI,thermo} and discussions around Eq. (\ref%
{purification})) for a time $\tau =1/(2T)$. However, this is not a
variational method, as the free energy does not necessarily decrease along
the path. Furthermore, one also looses the property that the desired result
is obtained as a fixed point (i.e. in the limit $\tau \rightarrow \infty $),
so that even though the thermal state may very well be represented by the
variational family, it is often not reached when the $\tau$-flow is finite.
In this paper we provide an alternative approach based on an equation
analogous to (\ref{Htau}) for mixed states, which ensures that the free
energy monotonically decreases and reaches a fixed point at precisely the
Gibbs state. We will demonstrate that this variational method can be applied
to minimize the free energy within a broad family of many-body states.

\begin{figure}[tbp]
\includegraphics[width=1.0\linewidth]{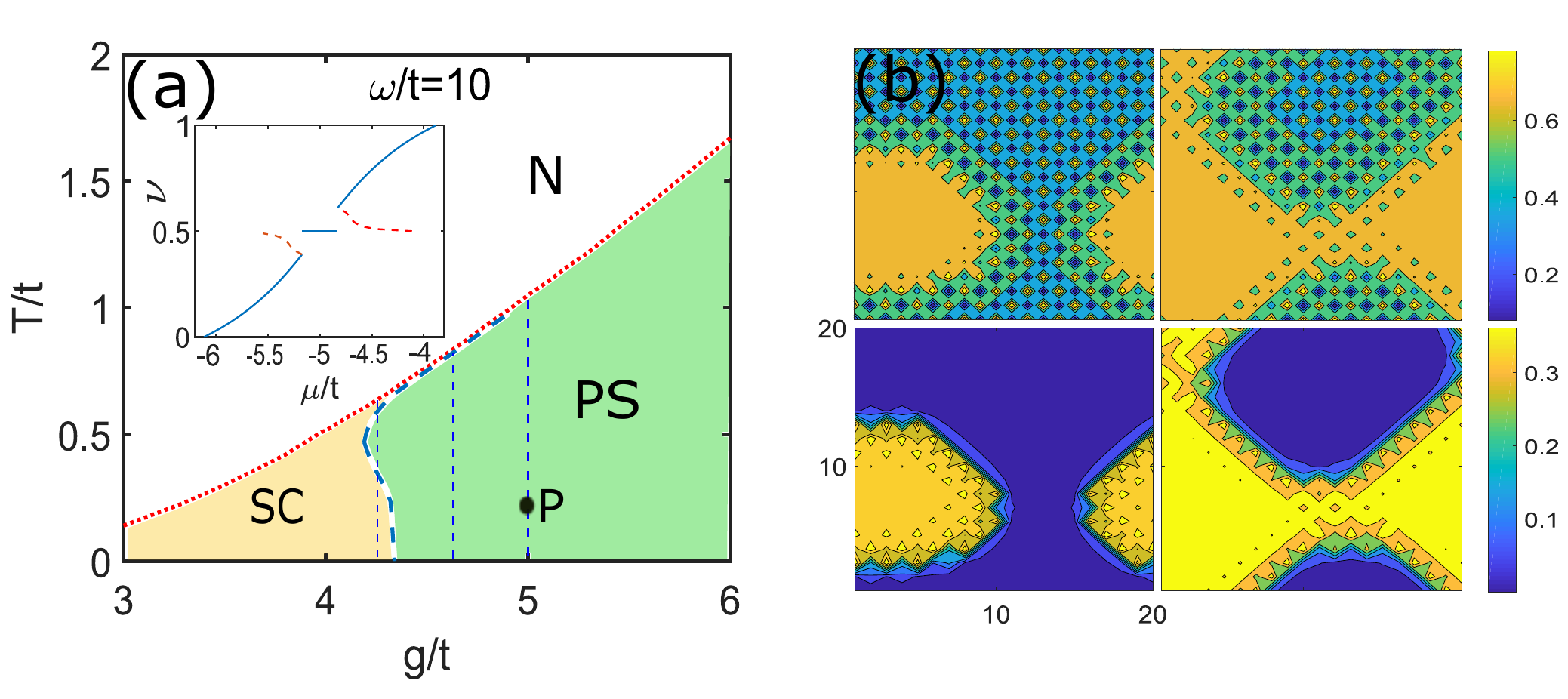}
\caption{(a) Phase diagrams for the Holstein model in a $50\times 50$
lattice for $\protect\omega_b/t=10$ and $\protect\nu=0.6$. The inset
displays the filling factor, $\protect\nu$, as a function of the chemical
potential at the point $P$, where $g/t=5$, $T/t=0.2$. (b) Phase separations
at $P$ for $\protect\nu=0.56$ (left panel) and $\protect\nu=0.6$ (right
panel) in a $20\times 20$ lattice. The first and second rows display the
electron density and the SC order parameter.}
\label{fig0}
\end{figure}

Regarding the choice of variational states, a particularly useful set for
the zero temperature case was introduced in \cite{Tao}, and consists of
states in the form
\begin{equation}
|\Psi (\xi )\rangle =U(\xi _{u})|\Psi _{G}(\xi _{g})\rangle .
\label{Varfamily}
\end{equation}%
Here, $\xi =(\xi _{u},\xi _{g})$ contains the variational parameters, $U\in
\mathcal{U}$, a set of unitary operators with a special form, and $\Psi _{G}$
is an arbitrary Gaussian state. The latter are those which can be written in
terms of a Gaussian function of creation and annihilation operators, and
they are very versatile as they can be fully characterized in terms of the
so-called covariance matrix and displacement vector (for the case of bosons)
\cite{QO,FG}. Furthermore, $U$ is non-Gaussian, so that it can encompass
different phenomena; in particular, in case one has both fermions and
bosons, it can describe non-trivial correlations among them, something which
is absent in Gaussian states. The TDVP method based on states of the form (%
\ref{Varfamily}) has been successfully applied to several problems. Those
include the Holstein and SSH models \cite{Tao}, polaron and spin-boson
problems \cite{Tao,Ultrastrong}, Kondo and Anderson impurity models \cite%
{Kondo,Anderson,Rydberg}, the 2D Hubbard-Holstein model \cite%
{Hubbard_Holstein} , and the Schwinger-like models with non-abelian gauge
groups \cite{LGT}.

In this Letter, we introduce a free energy flow based variational method
(FEFVM) to study systems at finite temperature, and show how it can be
applied to states of the form (\ref{Varfamily}). Firstly, we derive an
equation that extends parametric flow in (\ref{Htau}) to finite
temperatures, and which ensures that the free energy monotonically decreases
during the flow, so that, regardless of the initial density operator, the
system should ultimately flow to the Gibbs state. Secondly, we use a
purification of that state to re-express such flow equation in the form
similar to equation (\ref{Htau}). And finally, following \cite{Tao}, we show
how to apply this flow based formalism  to variational states of the form (%
\ref{Varfamily}), obtaining a set of differential equations for $\xi (\tau )$%
, which ensure that the free energy decreases during the flow. Thus, the
problem of studying finite temperature systems with FEFVM with such a family
of states possesses the same complexity as the standard ITVM for zero
temperature. We benchmark our method with the two dimensional (2D) negative-$%
U$ Hubbard model, for which standard mean-field theory can be easily
applied, and show that it yields more reliable results than the ITVM. Then,
we apply it to the 2D Holstein model, which describes electrons hopping on a
lattice and interacting with phonons. In Fig. \ref{fig0}a, we present the
resulting phase diagram for the phonon frequency $\omega _{b}/t=10$ and a
filling factor $\nu =0.6$, where $T $ is the temperature, $g$ the coupling
constant, and $t$ the hopping energy. As expected, our method predicts a
superconducting phase at low $g$ (when the model reduces to the BCS). For
higher values of $g$, it predicts separation between a auperconducting (SC)
and a charge-density wave (CDW) phases. This is obtained by both, a
homogeneous and a general variational ansatz. In the first case, this can be
deduced from the dependence of the filling factor on the chemical potential
(insert in Fig. \ref{fig0}a), whereas in the latter it explicitly follows
from the distribution of the electron density and the SC order parameter
(Fig. \ref{fig0}b). We point out that our approach predicts a CDW transition
temperature that monotonically increases with increasing electron-phonon
coupling strength. This temperature should be understood as the pseudogap
temperature of the onset of short-range correlations. Our method is
mean-field in its character and does not fully account for long-wavelength
fluctuations of the order parameter that determine the actual Tc. We expect
however that it accurately describes the increasing temperature of phase
separation.

\emph{Imaginary time evolution for the Free energy:} Given a Hamiltonian, $H$%
, and a temperature, $T$, we are interested in the Gibbs state described in
terms of the density operator%
\begin{equation}
\rho _{T}=\frac{e^{-\beta H}}{Z},  \label{Gibbs}
\end{equation}%
where $Z=\mathrm{tr}(e^{-\beta H})$ is the partition function, and $\beta
=1/T$. A unique feature of such an operator is that it minimizes the free
energy function%
\begin{equation}
f(\rho )=\mathrm{tr}(H\rho )-TS(\rho ),  \label{f}
\end{equation}%
where $S(\rho )=-\mathrm{tr}[\rho \ln (\rho )]$ is the von Neumann entropy
of $\rho $. The minimum of $f$ with respect to all possible density
operators is reached for $\rho =\rho _{T}$, so that this provides us with
the variational principle to determine the Gibbs state. Here, we will show
how this minimization can be done through a differential equation, akin to
the zero temperature state.

We define the free energy operator $F(\rho )=H+T\ln \rho $, so that $f(\rho
)=\mathrm{tr}[\rho F(\rho )]$. Now, let us consider the following equation%
\begin{equation}
d_{\tau }\rho =-\{F(\rho )-f(\rho ),\rho \}.  \label{imagtrho}
\end{equation}%
We want to show that any initial (normalized) state, $\rho (0)$, flows to $%
\rho _{T}$. For that, we will show that $d_{\tau }f[\rho (\tau )]\leq 0$
with the equality holding only if $\rho =\rho _{T}$. From the definition of $%
f(\rho )$, we have $d_{\tau }f(\rho )=\mathrm{tr}(Fd_{\tau }\rho )+T\mathrm{%
tr}[\rho d_{\tau }\ln (\rho )]$. The last term vanishes, since $\mathrm{tr}%
[\rho d_{\tau }\ln (\rho )]=\mathrm{tr}\left[ \int_{0}^{1}due^{(1-u)\ln
(\rho )}d_{\tau }[\ln (\rho )]e^{u\ln (\rho )}\right] =\mathrm{tr}\left[
d_{\tau }e^{\ln (\rho )}\right] =\mathrm{tr}(d_{\tau }\rho )=0$ where we
have utilized that Eq. (\ref{imagtrho}) conserves the trace of $\rho $.
Using Eq. (\ref{imagtrho}) we obtain
\begin{equation}
d_{\tau }f(\rho )=-2\mathrm{tr}\left[ \rho X(\rho )^{2}\right] \leq 0,
\end{equation}%
where we have defined $X(\rho )=[F(\rho )-f(\rho )]$. The derivative
vanishes when $X(\rho )=0$ which leads to (\ref{Gibbs}) \cite{footnote2}.
Note that while we refer to (\ref{imagtrho}) as imaginary time flow, it does
not correspond to the analytic continuation of the real time evolution of
the density matrix, except for the case of zero temperature.

The last piece we need is to transform (\ref{imagtrho}) into an equation
analogous to (\ref{Htau}). For that, we employ a particular purification of $%
\rho $, $\Phi _{p}$ (also called thermal double \cite{thermo}). This is done
by adding for each (bosonic or fermionic) mode an auxiliary one so that%
\begin{equation}
|\Phi _{p}\rangle =(\sqrt{\rho }\otimes {\openone})|\Phi ^{+}\rangle
\label{purification}
\end{equation}%
where $\Phi ^{+}$ is a maximally entangled state between each mode and the
corresponding ancilla, and thus fulfills $\mathrm{tr}_{a}[|\Phi ^{+}\rangle
\langle \Phi ^{+}|)\propto {\openone}$ \cite{footnote3}. We can recover $%
\rho $ out of $\Phi _{p}$ by simply tracing out the ancillas, i.e., $\rho =%
\mathrm{tr}_{a}(|\Phi _{p}\rangle \langle \Phi _{p}|)$. It follows directly
from (\ref{imagtrho}) that%
\begin{equation}
d_{t}|\Phi _{p}\rangle =-[F_{p}(\Phi _{p})-f_{p}(\Phi _{p})]|\Phi _{p}\rangle
\label{imagPuri}
\end{equation}%
where $F_{p}(\Phi )=F(\rho )\otimes {\openone}$ and $f_{p}(\Phi )=f(\rho )$,
with $\rho =\mathrm{tr}_{a}(|\Phi \rangle \langle \Phi |)$. The similarity
of Eqs. (\ref{imagPuri}) and (\ref{Htau}) is apparent, although the operator
$F_{p}$ explicitly depends on the state $\Phi $ and only acts non-trivially
on the system (and not on the ancillas). Thus, the resulting equation is
non-linear.

\emph{Variational method:} We are interested in approximating the Gibbs
state (\ref{Gibbs}) using the family of states%
\begin{equation}
\rho _{v}(\xi )=U(\xi _{u})\rho _{G}(\xi _{g})U(\xi _{u})^{\dagger }.
\label{rv}
\end{equation}%
In equation (\ref{rv}) $\rho _{G}$ in (\ref{rv}) is an arbitrary Gaussian
mixed state parametrized by $\xi _{g}$ with $\mathrm{tr}(\rho _{G})=1$. $U$
is a unitary operator which entangles different degrees of freedom and
allows us to describe states that do not obey Wick's theorem of Gaussian
ensembles. We consider the same family of unitary operators $U\in \mathcal{U}
$ that has been defined in the zero temperature case in Ref. \cite{Tao},
including all the conditions imposed on such operators. We assume that the
number of variational parameters $\xi _{u}$ in $U$ scales polynomially with
the system size. With the goal of describing states in (\ref{rv}) we
consider states in the doubled space (physical + ancilla) of the form
\begin{equation}
|\Psi _{v}(\xi )\rangle =[U(\xi _{u})\otimes {\openone}]|\Psi _{G}(\xi
_{g})\rangle  \label{VS}
\end{equation}%
with a normalized pure Gaussian state $\Psi _{G}$ as the purification of $%
\rho _{v}(\xi )$, namely, $\rho _{v}(\xi )=\mathrm{tr}_{a}(|\Psi _{v}(\xi
)\rangle \langle \Psi _{v}(\xi )|)$ as long as the trace over the auxiliary
modes reproduces $\rho _{G}(\xi _{g})=\mathrm{tr}_{a}(|\Psi _{G}(\xi
_{g})\rangle \langle \Psi _{G}(\xi _{g})|)$ \footnote{%
One could also add another $U\in\mathcal{U}$ acting on the ancilla and
depending on other variational parameters, as well as use a more symmetrized
version of Eq. (\ref{imagPuri}), see \cite{SM}}.
Note that in constructing the Gaussian state $\Psi _{G} $ we start with the
maximally entangled state between the ancilla and physical degrees of
freedom and then apply a Gaussian operator \cite{Tao} that acts only on the
physical degrees of freedom. 
Starting from Eq. (\ref{imagPuri}) it is possible to derive a set of
equations for variational parameters characterizing the purification. In
\cite{SM} we present details of such derivation and provide a simple proof
that the free energy decreases in the course of parametric flow with $\tau$,
as long as states have been chosen to be normalized. 
The method consists of projecting Eq. (\ref{imagPuri}) onto the tangent
plane of the manifold (\ref{VS}), in essentially the same way as it has been
done for the zero temperature case. The set $\mathcal{U}$ should be chosen
so that this can be done efficiently. Furthermore, a special feature of the
chosen family of variational states is that the free energy operator, $F$,
can be efficiently computed, since $\ln \rho _{v}=U(\xi _{u})\ln [\rho
_{G}(\xi _{g})]U(\xi _{v})^{\dagger }$, and the logarithm of a Gaussian
state can be readily calculated \cite{SM}.

As variational parameters for the Gaussian state, $\xi _{g}$, we use the
covariant matrix formalism. We consider a set of $N_{b}$ ($N_{f}$) bosonic
(fermionic) with annihilation operators $b_{n}$ ($c_{m}$). For a Gaussian
state we further define, as usual \cite{Tao,QO,FG,SM}, the covariance matrix
$\Gamma _{b,m}$ for the bosons and fermions, respectively, and the
displacement vector, $\Delta _{R}$ \cite{SM}.

Some conserved quantities $O$, e.g., the particle number $O=N$, may commute
with the many body Hamiltonian $H$. For the thermal state that breaks the
symmetry, i.e., $[\rho ,O]\neq 0$, we can fix the average value $%
\left\langle O\right\rangle $ in the flow equation by introducing a
time-dependent Lagrangian multiplier, which allows us to compute the
chemical potential \cite{SM}.

\emph{Negative-U Hubbard Model:} We first benchmark our method by analyzing
this textbook model, and show how it overcomes some of the deficiencies of
the ITVM. We consider the Hubbard Hamiltonian on a square lattice%
\begin{equation}
H_{\mathrm{BCS}}=-t\sum_{\left\langle nm\right\rangle ,\sigma }c_{n\sigma
}^{\dagger }c_{m\sigma }+U\sum_{n}c_{n\uparrow }^{\dagger }c_{n\downarrow
}^{\dagger }c_{n\downarrow }c_{n\uparrow },  \label{BCS}
\end{equation}%
where the first sum is restricted to nearest neighbors and $U<0$ describes
attractive interactions. This is a well known Hamiltonian, where BCS theory
correctly describes the appearance of a SC phase at sufficiently low
temperatures. The mean-field approach to the BCS model is known to be
quantitatively correct in the thermodynamic limit, although in one- and
two-dimensional systems the transition temperature should be understood as
that of opening of the quasi-particle gap, rather than the onset of the true
long range order.

We compare the results of our method with the mean-field calculation and the
ITVM mentioned in the introduction (see also \cite{SM}), and which is widely
used, for instance, in the context of matrix product states \cite{pMPS}. In
both, the ITVM and the FEFVM, we use the Fermionic Gaussian family of
translatinally invariant states (i.e., $(\Gamma _{m})_{{n,n^{\prime }}%
}=(\Gamma _{m})_{n-n^{\prime }}$). To account for the spontaneous symmetry
breaking in the SC phase, in the ITVM we introduce a small symmetry breaking
term in the Hamiltonian $\epsilon \sum_{n}c_{{n,\uparrow }}^{\dagger }c_{{%
n,\downarrow }}^{\dagger }$ and take $\epsilon \rightarrow 0$.

\begin{figure}[tbp]
\includegraphics[width=1.0\linewidth]{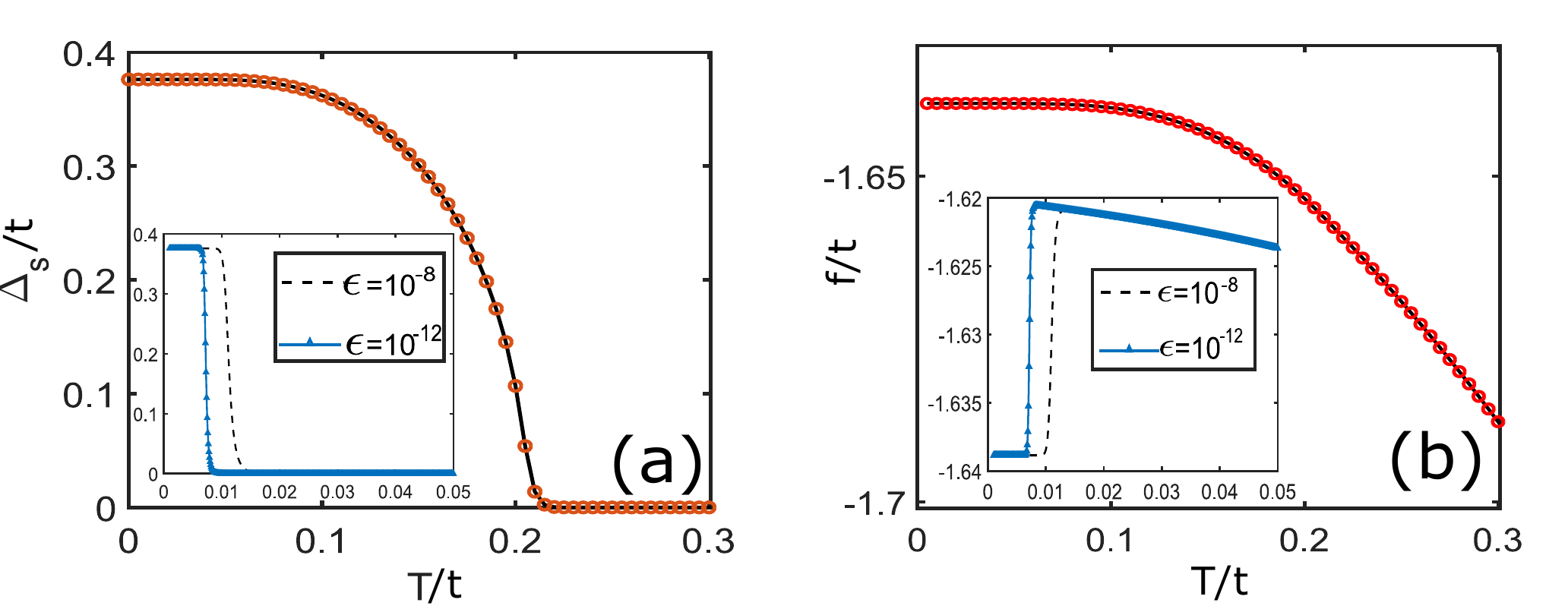}
\caption{Lattice BCS Model: $s $-wave order parameter (a) and free energy
density (b) as a function of the temperature $T/t$, for $U/t=-2$ and a $%
50\times 50$ lattice at half filling. The red curve gives the result of the
FEFVM, which is on top of the mean field result in the thermodynamic limit.
The black dashed line and the green curve (see insert) correspond to the
ITVM for a symmetry-breaking field with $\protect\epsilon=10^{-8},10^{-12}$,
respectively. }
\label{fig1}
\end{figure}

The results are displayed in Fig. \ref{fig1}, where we draw the $s$-wave
order parameter, $\Delta _{s}=U|\left\langle c_{n\downarrow }c_{n\uparrow
}\right\rangle |$, as a function of the temperature at half filling. The
figure shows that our method correctly reproduces the phase transition,
whereas the one based on ITVM does not. As mentioned above, the accumulation
of errors is responsible for this failure \cite{SM}.

\emph{Holstein Model: }We now investigate the 2D Holstein model, which
describes electrons on a lattice interacting with optical phonons. The
Hamiltonian is $H=H_{\mathrm{e}}+H_{\mathrm{ph}}+H_{\mathrm{int}}$, where $%
H_{\mathrm{e} }=-t\sum_{\left\langle n,m\right\rangle ,\sigma }c_{n\sigma
}^{\dagger }c_{m\sigma }$ and $H_{\mathrm{ph}}=\omega _{b}R^{T}R/4-\omega
_{b}/2$, where $t$ and $\omega_b$ are the electron hopping and phonon
frequency, respectively. The Holstein-type interaction $H_{\mathrm{int}%
}=g\sum_{n\sigma }x_{n}c_{n\sigma }^{\dagger }c_{n\sigma }$ between
electrons and phonons is characterized by the coupling strength $g$. For
weak electron phonon-interaction, $g\ll\omega_b$, one can eliminate the
bosons and obtain the Hubbard model so that, at sufficient low temperatures,
it displays an SC phase. For strong interactions and classical phonons,
Esterlis \textit{et al.} \cite{Esterlis} have used a Monte-Carlo analysis to
predict a commensurate CDW behavior that can be understood as the localized
phase of bipolarons.

We use the variational Ansatz (\ref{VS}) with the generalized Lang-Firsov
transformation $U=e^{S}$ \cite{Tao,LangFirsov}, where the generating
function $S=i\sum_{ln,\sigma }\lambda _{ln}p_{l}c_{n\sigma }^{\dagger
}c_{n\sigma }$ contains the variational parameters $\lambda _{ln}$. We use
two different kinds of Ansaetze for $\Delta _{R}$, $\Gamma _{b,m}$, and $%
\lambda _{ln}$:

\begin{description}
\item[(i)] General, where all components of the vector $\Delta_R$ and
matrices $\Gamma_b$, $\Gamma_m$, and $\lambda$ can take arbitrary values;

\item[(ii)] Homogeneous, where $\Delta _{R,l}=\Delta _{R,0}+(-1)^{l}\Delta
_{R,\pi }$ and $\xi _{n,n^{\prime }}=\xi _{0,n-n^{\prime }}+(-1)^{n}\xi
_{\pi ,n-n^{\prime }}$, with $\xi =\Gamma _{b},\Gamma _{m},\lambda $. Note
that in this way we can describe not only states with translational
symmetry, but also with CDW orders.
\end{description}

In both cases, the equations for the variational parameters can be easily
established \cite{SM} starting from (\ref{imagPuri}).

\begin{figure}[tbp]
\includegraphics[width=1.0\linewidth]{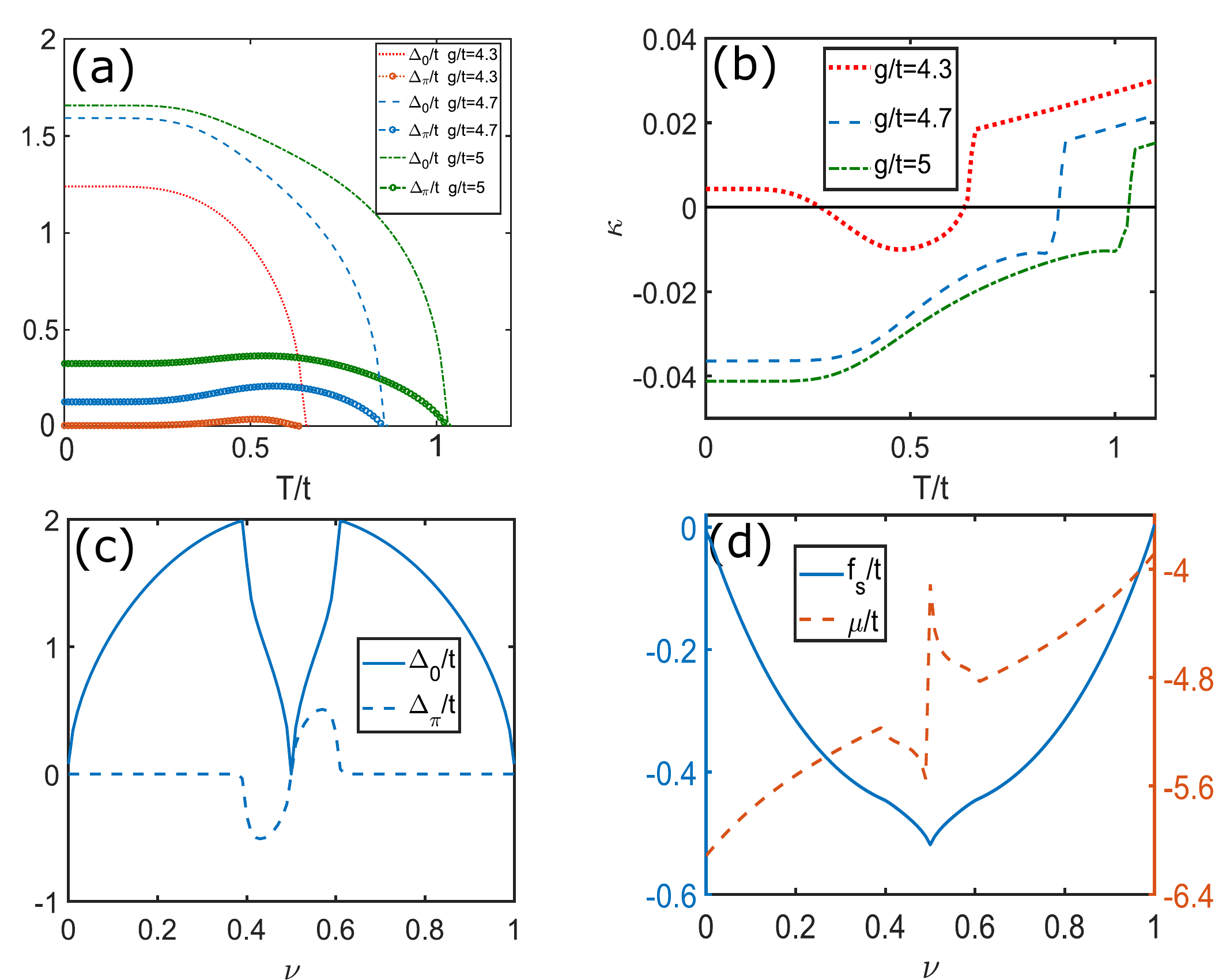}
\caption{(a)-(b) The SC order parameters and the compressibility along three
vertical lines in Fig. 1 obtained using homogeneous Ansaetze. Negative
compressibility indicates thermodynamically unstable states and corresponds
to phase separation. (c)-(d) The SC order parameters, the free energy, and
the chemical potential versus the filling factor $\protect\nu$ for $\protect%
\omega_b/t=10$, $g/t=5$, $T/t=0.2$. All plots have been obtained with the
homogeneous Ansatz. }
\label{fig2}
\end{figure}

In Fig. \ref{fig0}a, we show the phase diagram for the system with filling
factor $\nu=0.6$ and $\omega _{b}/t=10$. As expected, for relatively small $g
$ and low temperatures we find a SC phase. As $g$ increases, our method
predicts phase separation between SC and a CDW phase. While the naive Ansatz
(ii) predicts a supersolid phase (with non-vanishing SC and CDW order
parameters), the general Ansatz (i) establishes phase separation, as it is
shown in the snap-shots of Fig. \ref{fig0}b. One can recover this later
behavior from (ii) as well by computing the chemical potential as a function
of the filling factor $\nu $ (see insert in Fig. \ref{fig0}a). For $\nu$ in
the interval $\sim [0.35,0.65]$ this analysis predicts phase separation
between a CDW phase at half-filling, and a SC phase. The same result follows
from Maxwell construction \cite{MC}, in which one plots the free energy as a
function of $\nu $ and draws straight lines that are tangent to the free
energy $f$ and go through the minimum of $f$ (which occurs at half-filling).
Maxwell construction allows to predict the fractions of CDW and SC phases
for each value of the filling factor, $\nu $.

In Fig. \ref{fig2}a-b, we plot the order parameters $\Delta _{k=0,\pi
}=\sum_{n}e^{-ikn}V_{nn}\left\langle c_{n\downarrow }c_{n\uparrow
}\right\rangle /N$ and the compressibility $\kappa =\partial _{\nu }\mu $
for $g/t=4.3$, $4.7$, and $5$ as a function of the temperature (see also the
three vertical lines in \ref{fig0}a), where $V_{nn}=2(\omega
_{b}\sum_{l}\lambda _{ln}^{2}-2g\lambda _{nn})$ \cite{SM}. A negative value
indicates the onset of phase separation, which agrees with the corresponding
region of phase diagram of Fig. \ref{fig0}. To carry out the Maxwell
construction, in Fig. \ref{fig2}c-d, we display the order parameters $\Delta
_{0,\pi }$, the free energy $f_{s}=f+4g^{2}\nu /\omega _{b}$ (extracting the
phonon energy) and the chemical potential $\mu $ for $g/t=5$ and $T/t=0.2$
in a $50\times 50$ lattice. We have verified that this construction
reproduces the results of the full variational Ansatz (i).

\emph{Spectral functions:} Once we have obtained the variational
approximation to the Gibbs state, we can also consider the evolution of the
variational state in real time, which makes it possible to compute dynamical
response functions or even analyze pump and probe experiments. Generally,
one can consider situations when all parameters of the variational state in (%
\ref{rv}) become time dependent. However, when analyzing linear response it
is often sufficient to keep parameters of the unitary transformation to be
the same as in the equilibrium state and restrict the form of $\rho
_{G}=U(\xi _{g})\rho _{0}U(\xi _{g})^{\dagger }$, so that the evolution does
not change the spectrum of $\rho _{v}$. As one example of the dynamical
response function, the electron spectral function measured in ARPES
experiments can be calculated by extending the method reported in \cite%
{Anderson} to finite temperature (see SM4).

\emph{Conclusions: }We have developed a non-Gaussian variational approach to
minimize the free energy of many-body systems at finite temperature. We have
benchmarked it with the BCS and Holstein models. The later displays a
transition between the SC phase for weak coupling and phase separated regime
for stronger coupling. We find phase separation between the commensurate CDW
at half filling and a SC phase with either lower or higher density,
depending on whether the average density is below or above half-filling. Our
findings are consistent with the results obtained by the Monte-Carlo
analysis in the model with classical phonons \cite{Esterlis}. Formalism
developed in this paper can be extended to study broader classes of
electron-phonon models, including the Migdal-Eliashberg regime with $%
\omega_b< t$, systems with both electron-electron and electron-phonon
interactions, and systems with disorder.

\emph{Acknowledgements: } We thank I. Esterlis and Y. Wang for stimulating
discussions. T. S. acknowledges the Thousand-Youth-Talent Program of China.
J.I.C acknowledges the ERC Advanced Grant QENOCOBA under the EU Horizon2020
program (grant agreement 742102) and the German Research Foundation (DFG)
under Germany's Excellence Strategy through Project No. EXC-2111-390814868
(MCQST) and within the D-A-CH Lead-Agency Agreement through project No.
414325145 (BEYOND C). ED acknowledges support from the Harvard-MIT CUA,
Harvard-MPQ Center, AFOSR-MURI: Photonic Quantum Matter (award
FA95501610323), and DARPA DRINQS program (award D18AC00014).

\newpage \widetext

\begin{center}
\textbf{\large Supplemental Material}
\end{center}


\setcounter{equation}{0} \setcounter{figure}{0} 
\makeatletter

\renewcommand{\thefigure}{SM\arabic{figure}} \renewcommand{\thesection}{SM%
\arabic{section}} \renewcommand{\theequation}{SM\arabic{equation}}

This supplemental material is divided into five sections. In Sec. SM1, we
prove that the free energy decreases monotonically in the variational
manifold as well. In Sec. SM2, we recall the definition of quadratures and
covariance matrices for Gaussian states, and derive the explicit relation
between the Gaussian thermal state and the corresponding covariance
matrices. In Sec. SM3 we introduce a method to fix the expectation value of
any operator $O$ commuting with the Hamiltonian in the flow equation. In
Sec. SM4, we review the conventional purification methods to describe the
time evolution in real and imaginary time. The first give rise to the ITVM
mentioned in the text. As an application of the first, we give a technique
to compute spectral functions. In Sec. SM5, for the Holstein model, we
derove the equations of motion (EOM) for the parameters in the variational
state, including a generalized Lang-Firsov transformation.

\section{SM1. Monotonicity of the Free Energy in the FEFVM}

In this section we show that the evolution equations for the variational
parameters, $\xi =\{\xi _{j}\}$, ensure that the free energy decreases
monotonically with time so long as the states $\Psi (\xi )$ in the
variational family are normalized, that is, if we choose%
\begin{equation}
\langle \Psi (\xi )|\Psi (\xi )\rangle =1  \label{normaliz}
\end{equation}%
for all values of $\xi $. For that, let us first write the equation for the
variational state as
\begin{equation}
d_{\tau } |\Psi (\xi )\rangle= -\mathbb{P}[\xi (\tau )] \left( F\{\rho
\lbrack \xi (\tau )]\}\otimes \mathbb{I}-f\{\rho \lbrack \xi (\tau )]\}%
\mathbb{I}\otimes\mathbb{I}\right) |\Psi [\xi(\tau)]\rangle.  \label{dtauPsi}
\end{equation}%
In the following, in order to simplify the notation, we will not write
explicitly the dependence of the states and operators on $\xi $, the tensor
product, nor the identity operators. In (\ref{dtauPsi}), $\rho =\mathrm{tr}%
_{a}(|\Psi \rangle \langle \Psi |)$ is the reduced state, $f(\rho )$ defined
in Eq. (4), $F(\rho )=H+T\ln \rho $, and $\mathbb{P}$ is the projector onto
the tangential subspace spanned by $\partial _{\xi _{j}}|\Psi (\xi )\rangle$.

The normalization condition (\ref{normaliz}) implies
\begin{equation}
0=d_{\tau }\langle \Psi |\Psi \rangle =\langle (F\mathbb{P}+\mathbb{P}%
F)\rangle -2f\langle \mathbb{P}\rangle  \label{normcond}
\end{equation}%
where we have written $\langle \ldots \rangle =\langle \Psi |\ldots |\Psi
\rangle $. Thus, we have
\begin{equation}
d_{\tau }f=d_{\tau }\langle F\rangle =f\langle (F\mathbb{P}+\mathbb{P}%
F)\rangle -2\langle F\mathbb{P}F\rangle =-2\langle (F-f)\mathbb{P}%
(F-f)\rangle \leq 0
\end{equation}%
where we have used (\ref{normcond}) and the fact that for any operator $X$, $%
X\mathbb{P}X^{\dagger }$ is positive semi-definite. We have also utilized
that%
\begin{equation}
\langle d_{\tau }F\rangle =\mathrm{tr}(d_{\tau }\rho )=0.
\end{equation}%
Therefore, as announced, the free energy of the variational state decreases
under the FEFVM.

\section{SM2. Gaussian thermal states}

We define quadrature (Majorana) operators $x_{n}=b_{n}+b_{n}^{\dagger }$, $%
p_{n}=i(b_{n}^{\dagger }-b_{n})$ for the bosons [$a_{1,n}=c_{n}+c_{n}^{%
\dagger }$, $a_{2,n}=i(c_{n}^{\dagger }-c_{n})$ for the fermions]. We
collect these operators in column vectors $R=(x_{1},\ldots ,p_{1},\ldots
)^{T}$ and $A=(a_{1,1},\ldots ,a_{2,1},\ldots )^{T}$. The Gaussian state is
characterized by the quadrature and covariance matrices
\begin{subequations}
\begin{eqnarray}
\Delta _{R} &=&\langle \Psi _{G}|R|\Psi _{G}\rangle , \\
\Gamma _{b} &=&\frac{1}{2}\langle \Psi _{G}|\{\tilde{R},\tilde{R}^{T}\}|\Psi
_{G}\rangle , \\
\Gamma _{m} &=&\frac{i}{2}\langle \Psi _{G}|[A,A^{T}]|\Psi _{G}\rangle ,
\end{eqnarray}%
where $\tilde{R}=R-\Delta _{R}$ is the fluctuation around the average value.

We parametrize the Gaussian density matrix $\rho _{G}=e^{-K}/Z$ by
\end{subequations}
\begin{eqnarray}
K &=&\frac{1}{4}\tilde{R}^{T}\Omega _{b}\tilde{R}+i\frac{1}{4}A^{T}\Omega
_{m}A  \notag \\
&=&\frac{1}{4}\tilde{R}^{T}\Omega _{b}\tilde{R}+\frac{1}{2}C^{\dagger
}\Omega _{f}C
\end{eqnarray}%
with the matrices $\Omega _{b}$ and $\Omega _{m}$ (or $\Omega
_{f}=iW^{\dagger }\Omega _{m}W/2$ in the Nambu basis $C=(c,c^{\dagger })^{T}$%
), where the partition function $Z=\mathrm{tr}(e^{-K})$. We introduce the
unitary operators $U_{K}$ that transforms $\tilde{R}$ and $C$ as $%
U_{K}^{\dagger }RU_{K}=\bar{S}_{b}R+\Delta _{R}$ and $U_{K}^{\dagger }CU_{K}=%
\bar{U}_{f}C$, where the symplectic matrix $\bar{S}_{b}$ and the unitary
matrix $\bar{U}_{f}$ diagonalize $\Omega _{b}$ and $\Omega _{f}$, i.e., $%
\bar{S}_{b}^{T}\Omega _{b}\bar{S}_{b}=D_{b}$ and $\bar{U}_{f}^{\dagger
}\Omega _{b}\bar{U}_{f}=E_{f}$.

By definition, the covariance matrices are%
\begin{equation}
\Gamma _{b}=\frac{1}{2}\langle \Psi _{G}|\{\tilde{R},\tilde{R}^{T}\}|\Psi
_{G}\rangle =\bar{S}_{b}\frac{1}{2}tr(\bar{\rho}_{G}\{R,R^{T}\})\bar{S}%
_{b}^{T}=\bar{S}_{b}\coth (\frac{D_{b}}{2})\bar{S}_{b}^{T},
\end{equation}%
and%
\begin{equation}
\Gamma _{f}=\langle \Psi _{G}|CC^{\dagger }|\Psi _{G}\rangle =\bar{U}_{f}%
\frac{1}{e^{-E_{f}}+1}\bar{U}_{f}^{\dagger }=\frac{1}{e^{-\Omega _{f}}+1},
\label{Gf}
\end{equation}%
where we have used the property that the density matrix $\bar{\rho}%
_{G}=U_{K}^{\dagger }\rho _{G}U_{K}$ describes the thermal state of free
bosons and fermions.

Using the symplectic property $\bar{S}_{b}\Sigma ^{y}\bar{S}_{b}^{T}=\Sigma
^{y}$ and the fact that $\coth (x/2)$ is an odd function, we can re-express $%
\Gamma _{b}$ in the compact form%
\begin{equation}
\Gamma _{b}=\frac{e^{\Sigma ^{y}\Omega _{b}}+1}{e^{\Sigma ^{y}\Omega _{b}}-1}%
\Sigma ^{y},  \label{Gb}
\end{equation}
where $\Sigma ^{y}=I_{N_{b}}\otimes \sigma ^{y}$ is determined by the Pauli
matrix $\sigma ^{y}$. By inverting Eqs. (\ref{Gf}) and (\ref{Gb}), we obtain%
\begin{equation}
\Omega _{b}=\Sigma ^{y}\ln \frac{\Gamma _{b}\Sigma ^{y}+1}{\Gamma _{b}\Sigma
^{y}-1},
\end{equation}%
and%
\begin{equation}
\Omega _{m}=i\ln (\frac{1+i\Gamma _{m}}{1-i\Gamma _{m}}).
\end{equation}

\section{SM3. Conserved quantities under the FEFVM}

For a system with a conserved quantity $O$, i.e., $[O,H]=0$, the thermal
state may break that symmetry, i.e., $[O,\rho ]\neq 0$. A typical example is
the $U(1)$ symmetry breaking in superconductors, where the total fermion
number operator $N$ commutes with the Hamiltonian, however, the thermal
state breaks that symmetry. In this section, we introduce a time-dependent
term in the flow equation to fix the average value $\left\langle
O\right\rangle $.

We modify the flow equation (8) to%
\begin{equation}
\partial _{\tau }\left\vert \Psi \right\rangle =-\mathbf{P}_{O}(F-\mathcal{E}%
)\left\vert \Psi \right\rangle ,  \label{dPhim}
\end{equation}%
where the projector%
\begin{equation}
\mathbf{P}_{O}=1-\frac{1}{\mathcal{N}_{O}}O\left\vert \Psi \right\rangle
\left\langle \Psi \right\vert O,
\end{equation}%
$\mathcal{N}_{O}=\left\langle \Psi \right\vert O^{2}\left\vert \Psi
\right\rangle $, and $\mathcal{E}$ is a time dependent function to be
determined. It immediately follows that Eq. (\ref{dPhim}) leads to the
conservation law $\partial _{\tau }\left\langle O\right\rangle =0$.

To fulfill the normalization condition $\partial _{\tau }\left\langle \Psi
\left\vert \Psi \right\rangle \right. =0$, we choose%
\begin{equation}
\mathcal{E}=\frac{\left\langle F\right\rangle -\frac{1}{\mathcal{N}_{O}}%
\left\langle O\right\rangle \left\langle OF\right\rangle }{1-\frac{1}{%
\mathcal{N}_{O}}\left\langle O\right\rangle ^{2}}.
\end{equation}%
A straightforward calculation results in
\begin{equation}
\partial _{\tau }\left\vert \Psi \right\rangle =-(\bar{F}-\left\langle \bar{F%
}\right\rangle )\left\vert \Psi \right\rangle ,
\end{equation}%
where the new free energy operator $\bar{F}=F-\mu _{O}O$ is modified, with a
time-dependent function
\begin{equation}
\mu _{O}=\frac{\left\langle OF\right\rangle -\left\langle O\right\rangle
\left\langle F\right\rangle }{\mathcal{N}_{O}-\left\langle O\right\rangle
^{2}},
\end{equation}

For $O=N$,the particle number, $\mu _{O}$ is the chemical potential, which
adjusts itself during the flow in order to keep the average value $%
\left\langle N\right\rangle $ unchanged. This equation can be projected onto
the tangent plane of the variational manifold in order to obtain the
differential equations for the variational parameters.

\section{SM4. Imaginary and real time evolutions through purification}

In the standard purification method, the thermal state $\rho _{T}=e^{-\beta
H}/Z$ can be written as $\rho _{T}=\mathrm{tr}_{a}|\Phi _{p}\rangle
\left\langle \Phi _{p}\right\vert $ with%
\begin{equation}
|\Phi _{p}\rangle =\frac{1}{\sqrt{Z}}(e^{-\frac{1}{2}\beta H}\otimes {%
\openone})|\Phi ^{+}\rangle ,
\end{equation}%
or, in a more symmetric form, $\rho _{T}=\mathrm{tr}_{a}|\Phi _{s}\rangle
\left\langle \Phi _{s}\right\vert $ with%
\begin{equation}
|\Phi _{s}\rangle =\frac{1}{\sqrt{Z}}(e^{-\frac{1}{4}\beta H}\otimes e^{-%
\frac{1}{4}\beta \bar{H}})|\Phi ^{+}\rangle ,
\end{equation}%
where $|\Phi ^{+}\rangle $ is the maximal entangled state.

For bosons, the Hamiltonian of the ancillas is $\bar{H}=H^{T}$. For
fermions, we notice the relation%
\begin{equation}
c|\Phi ^{+}\rangle =d^{\dagger }|\Phi ^{+}\rangle ,c^{\dagger }|\Phi
^{+}\rangle =-d|\Phi ^{+}\rangle
\end{equation}%
for the annihilation and creation operators of the system and ancillas
acting\ on $|\Phi ^{+}\rangle =(1+c^{\dagger }d^{\dagger })|0\rangle /\sqrt{2%
}$. As a result, one has to add a minus sign for the creation operator,
corresponding to a particle-hole transformation between system and ancilla.

Let us consider now the imaginary time evolution dictated by a Hamiltonian $H
$. The EOM for $|\Phi _{p}\rangle $ and $|\Phi _{s}\rangle $ are%
\begin{equation}
\partial _{\tau }|\Phi _{p}\rangle =-\frac{1}{2}(H\otimes {\openone}%
-\left\langle H\right\rangle )|\Phi _{p}\rangle  \label{dp}
\end{equation}%
and%
\begin{equation}
\partial _{\tau }|\Phi _{s}\rangle =-\frac{1}{4}(H\otimes {\openone+\openone}%
\otimes H-2\left\langle H\right\rangle )|\Phi _{s}\rangle .  \label{ds}
\end{equation}%
The thermal Gibbs state is obtained by evolving this state starting from $%
\Phi^+$ for a time $\tau=\beta/2$. One can project this equation onto the
tangent plane of any variational manifold in order to obtain a practicable
method to study thermal equilibrium, which leads to the ITVM as described in
the main text.

The standard purification method works very well if the solutions $|\Phi
_{p,s}\rangle $ are exact; however, it may not give reliable results for
variational states. One can track the reason for the potential failure of
this method as follows. First, the standard method accumulates error along
the time evolution up to the time $\tau =\beta /2$. The FEFVM, however,
obtains the purified state at a fixed point, $\tau \rightarrow \infty $, and
thus it does not depend on the path used to reach it. For the ITVM, this can
be seen very clearly as follows in the BCS model described in the main text.
Since we are dealing with Gaussian states, the projection onto the tangent
plane can be translated into a differential equation of the form $d_{\tau
}|\Psi _{G}\rangle =-H_{\mathrm{P}}(\tau )|\Psi _{G}\rangle $, where the
projected Hamiltonian depends on the variational parameters and is thus time
dependent. The solution to this equation can be written as $|\Psi (\tau
)\rangle \propto \mathcal{T}\exp [-\int_{0}^{\tau }d\tau ^{\prime }H_{%
\mathrm{P}}(\tau ^{\prime })]|\Phi ^{+}\rangle $, whereas we know that state
that minimizes the free energy must have a purification of the form $|\Psi
(\tau )\rangle \propto \exp [-H_{\mathrm{eff}}(\tau )]|\Phi ^{+}\rangle $.
Furthermore, the ITVM does not perform well whenever there is symmetry
breaking. Since the initial thermal state with infinite temperature
maintains all the symmetries, one has to add a small symmetry breaking term
in the Hamiltonian. However, the appearance of the symmetry breaking is very
sensitive to that term, and the corresponding order parameter only agrees
with that from the correct BCS theory near zero temperature.

Let us now move to the variational study of real time evolution of mixed
states. In this case, the density matrix $\rho $ obeys the Liouville equation%
\begin{equation}
i\partial _{t}\rho =[H,\rho ],  \label{LE}
\end{equation}%
where in general $\rho $ can be a mixed state. We introduce the purification
for $\rho =\mathrm{tr}_{a}|\Phi _{p}\rangle \left\langle \Phi
_{p}\right\vert $, where%
\begin{equation}
|\Phi _{p}\rangle =(\sqrt{\rho }\otimes {\openone})|\Phi ^{+}\rangle .
\end{equation}

One can easily show that the Schr\"{o}dinger equation%
\begin{equation}
i\partial _{t}|\Phi _{p}\rangle =H\otimes {\openone}|\Phi _{p}\rangle
\label{SEp}
\end{equation}%
leads to Eq. (\ref{LE}). The Eq. (\ref{SEp}) can then be solved
variationally \cite{Tao}.

The purified Schr\"{o}dinger Eq. (\ref{SEp}) can be applied to study the
spectral function $A(\omega )=-$Im$G_{R}(\omega )/\pi $, where $G_{R}(\omega
)$ is the Fourier transformation of the retarded Green function%
\begin{equation}
G_{R}(t)=-itr\rho \{c(t),c^{\dagger }\}\theta (t)
\end{equation}%
defined in some basis $c=(c_{1},c_{2},...,c_{N})^{T}$. By the purification,
the Green function becomes%
\begin{eqnarray}
G_{R}(t) &=&-i\left\langle \Phi _{p}\right\vert \{c(t),c^{\dagger }\}|\Phi
_{p}\rangle \theta (t)  \notag \\
&=&-i\left\langle \Phi _{p}\right\vert e^{iHt}ce^{-iHt}c^{\dagger }|\Phi
_{p}\rangle \theta (t)-i\left\langle \Phi _{p}\right\vert c^{\dagger
}e^{iHt}ce^{-iHt}|\Phi _{p}\rangle \theta (t).  \label{GR}
\end{eqnarray}

By taking the second term in Eq. (\ref{GR}) as an example, we first
calculate the real-time evolution $|\Phi _{p}(t)\rangle =e^{-iHt}|\Phi
_{p}\rangle $ by using Eq. (\ref{SEp}). The second real-time evolution $|%
\bar{\Phi}_{p}(t)\rangle =e^{iHt}(c|\Phi _{p}(t)\rangle )$ can also be
obtain by solving Eq. (\ref{SEp}), where the Hamiltonian $H$ is replaced by $%
-H$. Finally, the second term in Eq. (\ref{GR}) becomes the overlap $%
-i\left\langle \Phi _{p}\right\vert c^{\dagger }|\bar{\Phi}_{p}(t)\rangle $.

In practice, one has to carefully choose the variational manifold $\mathcal{M%
}$, such that $\{|\Phi _{p}\rangle $, $c|\Phi _{p}(t)\rangle $, $|\bar{\Phi}%
_{p}(t)\rangle \}\in \mathcal{M}$ and the overlap $-i\left\langle \Phi
_{p}\right\vert c^{\dagger }|\bar{\Phi}_{p}(t)\rangle $ can be evaluated
efficiently. The Gaussian variational manifold satisfies this condition,
where Eq. (\ref{SEp}) is projected in the Gaussian manifold. It is worthy to
remark that the time-dependent global phase in the real time evolution is
crucial for the spectral function, which can be tracked by the Wei-Norman
algebra method \cite{Tao}.

A further approximation can be applied to simplify the calculation of $%
A(\omega )$. The Hamiltonian $H$ in Eq. (\ref{GR}) can be approximated by
the mean-field Hamiltonian $H_{\mathrm{MF}}=C^{\dagger }\mathcal{H}_{f}C/2$
in the Nambu basis $C=(c,c^{\dagger })$ \cite{Anderson}, where $\mathcal{H}%
_{f}$ is constructed in the equilibrium state by the Wick theorem, similarly
to the treatment in the superconductivity theory. As a result, the Green
function $G_{R}(t)=-ie^{-i\mathcal{H}_{f}t}\theta (t)$, and the spectral
function $A(\omega )=\delta (\omega -\mathcal{H}_{f})$ displays the peaks
corresponding to the quasi-particle energy. In the electron-phonon
interacting system, the phonon broadening effects in the spectral function
can be included by the expansion of $H=H_{\mathrm{MF}}+H_{I}$ in the
vicinity of the Gaussian thermal state $\rho $, where the mean-field
Hamiltonian $H_{\mathrm{MF}}$ has the quadratic form and $H_{I}$ contains
the higher order terms. The perturbation theory gives rise to the
renormalization of the quasi-particle energy and the broadening of the peak
in $A(\omega )$.

\section{SM5. Application to Holstein models}

We derive the EOM of $\Delta _{R}$, $\Gamma _{b,m}$, and $\lambda _{ln}$ for
the Holstein model by projecting Eq. (8) on the tangential space. The
tangential vextor of the variational ansatz (10) determined by the
Lang-Firsov transformation $U=e^{S}$ reads%
\begin{equation}
d_{\tau }|\Psi _{v}\rangle =e^{S}[i\sum_{ln,\sigma }d_{\tau }\lambda
_{ln}p_{l}c_{n\sigma }^{\dagger }c_{n\sigma }|\Psi _{G}\rangle +d_{\tau
}|\Psi _{G}\rangle ]
\end{equation}%
The Gaussian state $|\Psi _{G}\rangle =U_{\mathrm{GS}}|0\rangle $ is
determined by the unitary operator $U_{\mathrm{GS}}$ that transforms $C$ and
$R$ as $U_{\mathrm{GS}}^{\dagger }RU_{\mathrm{GS}}=\Delta _{R}+S_{b}R$ and $%
U_{\mathrm{GS}}^{\dagger }CU_{\mathrm{GS}}=U_{f}C$, where $S_{b}$ and $U_{f}$
are the time-dependent symplectic and unitary matrices.

The time derivative to $|\Psi _{G}\rangle $ gives rise to the tangential
vector $d_{\tau }|\Psi _{v}\rangle =e^{S}U_{\mathrm{GS}}(|V_{1}\rangle
+|V_{2}\rangle +|V_{3}\rangle )$. The tangential vector $|V_{1}\rangle
=-R^{T}S^{T}\xi _{1}|0\rangle /2$ containing the linear operator is
determined by%
\begin{equation}
\xi _{1}=\Sigma ^{y}d_{\tau }\Delta _{R}-2i\sum_{n\sigma }d_{\tau }\lambda
_{ln}\left\langle c_{n\sigma }^{\dagger }c_{n\sigma }\right\rangle .
\end{equation}%
The tangential vector%
\begin{equation}
|V_{2}\rangle =\text{:}[-\frac{1}{4}R^{T}S_{b}^{T}\Sigma ^{y}d_{\tau }S_{b}R+%
\frac{1}{2}C^{\dagger }U_{f}^{\dagger }(d_{\tau }+O_{f})U_{f}C]\text{:}%
|0\rangle
\end{equation}%
contains the quadratic normal ordered operators acting on the vacuum state,
where the anti-Hermitian matrix $O_{f}=\sigma ^{z}\otimes I_{2}\otimes
diag(i\sum_{l}\left\langle p_{l}\right\rangle d_{\tau }\lambda _{ln})$. The
tangential vector%
\begin{equation}
|V_{3}\rangle =i\sum_{ln,\sigma }(R^{T}S^{T})_{l}d_{\tau }\lambda _{ln}\text{%
:}U_{\mathrm{GS}}^{\dagger }c_{n\sigma }^{\dagger }c_{n\sigma }U_{\mathrm{GS}%
}\text{:}|0\rangle
\end{equation}%
contains the cubic normal ordered operators.

The right hand side of Eq. (8) can be written as%
\begin{equation}
|V_{R}\rangle =-e^{S}[\bar{H}+T\ln \rho _{G}-f(\rho )]U_{\mathrm{GS}%
}|0\rangle ,
\end{equation}%
where%
\begin{eqnarray}
\bar{H} &=&-t\sum_{\left\langle nm\right\rangle ,\sigma
}e^{-i\sum_{l}p_{l}(\lambda _{ln}-\lambda _{lm})}c_{n\sigma }^{\dagger
}c_{m\sigma }-\mu \sum_{n\sigma }c_{n\sigma }^{\dagger }c_{n\sigma }+\frac{1%
}{4}\omega _{b}R^{T}R-\frac{1}{2}\omega _{b}  \notag \\
&&+\sum_{ln,\sigma }R_{l}^{T}G_{ln}c_{n\sigma }^{\dagger }c_{n\sigma }+\frac{%
1}{2}\sum_{n\sigma ,m\sigma ^{\prime }}V_{nm}c_{n\sigma }^{\dagger
}c_{n\sigma }c_{m\sigma ^{\prime }}^{\dagger }c_{m\sigma ^{\prime }}
\end{eqnarray}%
is determined by the renormalized electron-phonon interaction $G_{ln}=\left(
g\delta _{ln}-\omega _{b}\lambda _{ln},0\right) ^{T}$ and the
electron-electron interaction%
\begin{equation}
V_{nm}=2[\omega _{b}\sum_{l}\lambda _{ln}\lambda _{lm}-g(\lambda
_{nm}+\lambda _{mn})]
\end{equation}%
induced by mediating phonons.

In the state $|V_{R}\rangle $, we can move the Gaussian unitary operator $U_{%
\mathrm{GS}}$ to the left side of the free energy operator, and obtain the
vector $|V_{R}\rangle =-e^{S}U_{\mathrm{GS}}(|R_{1}\rangle +|R_{2}\rangle
+|R_{3}\rangle )$. The linear vector reads%
\begin{eqnarray}
|R_{1}\rangle &=&[\frac{1}{2}\omega _{b}R^{T}S^{T}\Delta
_{R}+\sum_{ln,\sigma }(R^{T}S^{T})_{l}G_{ln}\left\langle c_{n\sigma
}^{\dagger }c_{n\sigma }\right\rangle  \notag \\
&&+i\sum_{\left\langle nm\right\rangle ,\sigma }t_{\mathrm{eff}%
,nm}(R^{T}S^{T})_{l}(\lambda _{ln}-\lambda _{lm})\left\langle c_{n\sigma
}^{\dagger }c_{m\sigma }\right\rangle ]|0\rangle ,
\end{eqnarray}%
where the renormalized hopping strength%
\begin{equation}
t_{\mathrm{eff},nm}=te^{-i\sum_{l}\left\langle p\right\rangle _{l}(\lambda
_{ln}-\lambda _{lm})}e^{-\frac{1}{2}w_{l,nm}\Gamma _{b,ll^{\prime
}}w_{l^{\prime },nm}},
\end{equation}%
and $w_{l,nm}=\lambda _{ln}-\lambda _{lm}$.

The quadratic vector%
\begin{equation}
|R_{2}\rangle =\frac{1}{4}R^{T}S^{T}\Omega _{\mathrm{MF}}SR+\frac{1}{2}%
C^{\dagger }U_{f}^{\dagger }\mathcal{F}_{f}U_{f}C
\end{equation}%
contains the bosonic and fermionic parts. The mean-field free energy $\Omega
_{\mathrm{MF}}=\Omega _{0}-T\Omega _{b}$ of phonons is determined by%
\begin{equation}
\Omega _{0}=\omega _{b}+2\sum_{\left\langle nm\right\rangle ,\sigma }t_{%
\mathrm{eff},nm}\left\langle c_{n\sigma }^{\dagger }c_{m\sigma
}\right\rangle w_{l,nm}w_{l^{\prime },nm}.
\end{equation}%
The mean-field free energy in the Dirac basis is $\mathcal{F}_{f}=\mathcal{H}%
_{f}-iW^{\dagger }T\Omega _{m}W/2$:%
\begin{equation}
\mathcal{H}_{f}=\left(
\begin{array}{cc}
\mathcal{E} & \Delta \\
\Delta ^{\dagger } & -\mathcal{E}^{T}%
\end{array}%
\right) ,
\end{equation}%
where the dispersion relation and effective chemical potential%
\begin{eqnarray}
\mathcal{E} &=&-t_{\mathrm{eff},nm}-V_{nm}\left\langle c_{m\sigma ^{\prime
}}^{\dagger }c_{n\sigma }\right\rangle -\mu _{\mathrm{eff,}n\sigma }, \\
\mu _{\mathrm{eff,}n\sigma } &=&\mu -(\frac{1}{2}V_{nn}+\sum_{l}\Delta
_{R,l}^{T}G_{ln})-\sum_{m\sigma ^{\prime }}V_{nm}\left\langle c_{m\sigma
^{\prime }}^{\dagger }c_{m\sigma ^{\prime }}\right\rangle
\end{eqnarray}%
of the phonon dressed polaron contains the Hartree-Fock corrections, and the
gap matrix $\Delta _{n\sigma ,m\sigma ^{\prime }}=V_{nm}\left\langle
c_{m\sigma ^{\prime }}c_{n\sigma }\right\rangle $. The free energy of
electrons in the Majorana basis is related to $\mathcal{F}_{f}$ as $\mathcal{%
F}_{m}=-iW\mathcal{F}_{f}W^{\dagger }/2$. The cubic vector is%
\begin{equation}
|R_{3}\rangle =\sum_{l}(R^{T}S^{T})_{l}\text{:}U_{\mathrm{GS}}^{\dagger
}[\sum_{\left\langle nm\right\rangle ,\sigma }it_{\mathrm{eff}%
,nm}w_{l,nm}c_{n\sigma }^{\dagger }c_{m\sigma }+\sum_{n\sigma
}G_{ln}c_{n\sigma }^{\dagger }c_{n\sigma }]U_{\mathrm{GS}}\text{:}.
\end{equation}

The projection on the linear tangential vector gives rise to the EOM%
\begin{eqnarray}
d_{\tau }\Delta _{R} &=&-\Gamma _{b}[\omega _{b}\Delta _{R}+2\sum_{n\sigma
}G_{ln}\left\langle c_{n\sigma }^{\dagger }c_{n\sigma }\right\rangle  \notag
\\
&&+2i\sum_{\left\langle nm\right\rangle ,\sigma }t_{\mathrm{eff},nm}(\lambda
_{ln}-\lambda _{lm})\left\langle c_{n\sigma }^{\dagger }c_{m\sigma
}\right\rangle ]  \notag \\
&&+2i\Sigma ^{y}\sum_{n\sigma }d_{\tau }\lambda _{ln}\left\langle c_{n\sigma
}^{\dagger }c_{n\sigma }\right\rangle  \label{dR}
\end{eqnarray}%
for the quadrature $\Delta _{R}$. The projection on the quadratic tangential
vector results in the EOM%
\begin{eqnarray}
d_{\tau }\Gamma _{b} &=&\Sigma ^{y}\Omega _{\mathrm{MF}}\Sigma ^{y}-\Gamma
_{b}\Omega _{\mathrm{MF}}\Gamma _{b},  \label{dGb} \\
d_{\tau }\Gamma _{f} &=&\{\Gamma _{f},\mathcal{F}_{f}\}-2\Gamma _{f}\mathcal{%
F}_{f}\Gamma _{f}+[\Gamma _{f},O_{f}]  \label{dGf}
\end{eqnarray}%
for the covariance matrices $\Gamma _{b}$ and $\Gamma _{f}=1/2-iW^{\dagger
}\Gamma _{m}W/4$.

The projection on the cubic tangential vector leads to%
\begin{equation}
d_{\tau }\lambda _{ln}=\sum_{l^{\prime }}(\Gamma _{b,p}^{-1})_{ll^{\prime
}}G_{l^{\prime }n}+\sum_{m}v_{lm}D_{mn}^{-1},  \label{dL}
\end{equation}%
where $v_{lm}=\sum_{\delta ,\sigma }w_{l,m+\delta m}t_{\mathrm{eff}%
,mm+\delta }$Re$\left\langle c_{m\sigma }^{\dagger }c_{m+\delta \sigma
}\right\rangle $ and the connected correlation function%
\begin{equation}
D_{nm}=\sum_{\sigma \sigma ^{\prime }}\left\langle c_{n\sigma }^{\dagger
}c_{n\sigma }c_{m\sigma ^{\prime }}^{\dagger }c_{m\sigma ^{\prime
}}\right\rangle _{c}.
\end{equation}


\begin{thebibliography}{99}
\bibitem{footnote} Note that in principle this procedure may lead to the
flow stopping at a local rather than the global minimum. However, the
experience so far shows that this can be overcome by taking different
initial values for $\xi (0)$.
\bibitem{footnote2} Strictly speaking, this is the only solution as long as
we impose that $\rho $ has full rank.
\bibitem{footnote3} For the bosonic case, the state $\Phi ^{+}$ is not
normalizable. However, since we use the covariant matrix formalism, this
problem can be easily circumvented.
\end{thebibliography}
\end{document}